# Parallel Evolutionary Computation in Very Large Scale Eigenvalue Problems

Hesam T. Dashti[1], Alireza F. Siahpirani[2], Liya Wang[3], Mary Kloc[4], and Amir H. Assadi[5]
[1]Department of Computer Science, University of Tehran, Iran
[2]Department of Mathematics, Statistics, and Computer Science, University of Tehran, Iran
[3]Departments of Mathematics and Botany, University of Wisconsin, Madison, Wisconsin, USA
[4] Department of Applied Mathematics and Computer Science, Weizmann Institute of Science, Rehovot, Israel
[5]Department of Mathematics, University of Wisconsin, Madison, Wisconsin, USA

**Abstract** – *The history of research on eigenvalue problems is rich with many outstanding contributions. Nonetheless, the rapidly increasing size of data sets requires new algorithms for old problems in the context of extremely large matrix dimensions [21]. This paper reports on new methods for finding eigenvalues of very large matrices by a synthesis of evolutionary computation, parallel programming, and empirical stochastic search. The direct design of our method has the added advantage that it could be adapted to extend many algorithmic variants of solutions of generalized eigenvalue problems to improve the accuracy of our algorithms. The preliminary evaluation results are encouraging and demonstrate the method's efficiency and practicality.*

## 1 Introduction

Broadly speaking, the computation of eigenvalues and eigenvectors follows one of the two major approaches. In the first approach, direct deterministic algorithms provide precise solutions limited only by machine precision. The second approach covers heuristic methods, and it belongs to the family known as approximation algorithms, which are expected to converge to true solutions [4] [14]. The direct algorithms follow predetermined steps and, in specified time complexity, reach a solution that could be approximated up to arbitrary precision. Since the eigenvalues of a matrix are roots of its characteristic polynomial, and in view of the impossibility of finding direct algorithms for the exact calculation of the roots of polynomials of degree greater than four (for instance, by Galois theory), we must expect an iterative procedure as part of any eigenvalue-eigenvector computation. Thus, an algorithm for a general matrix (i.e. not a diagonal matrix, triangular matrix, etc.) is necessarily iterative and the problem is to identify iterative algorithms that have fast rates of convergence and lead to accurate results. The solution strategy for an eigenvalue problem depends also on a number of properties that greatly affect the choice of algorithm, such as real or complex entries or special properties that the matrix satisfies [4] [12] [14].

One of the earliest methods for approximating solutions of eigenvalue problems is the classical Matrix Power Method (MPM), which finds the first few dominant eigenvalues of large matrices [14]. The main idea in MPM-type methods is that for almost any vector X that is repeatedly transformed by the matrix M (and is properly normalized), eventually images of X will align in the direction of the eigenvector associated with the eigenvalues that are largest in absolute value. The convergence rate for MPM depends on the ratio of the second largest eigenvalue (in absolute value) to the largest eigenvalue (in absolute value), and for many applications this leads to unacceptably slow convergence [15] [16]. Also, MPM can be problematic if one wants to compute a number of extremal eigenvalues. Nonetheless, the MPM is still in use, for instance, as a part of Krylov methods, inverse iteration, and QR-method [16] [17] [18] [19].

The naïve implementations of MPM and many other algorithms developed in the 20[th] century have limitations in their accuracy and rate of convergence with increasing size dimensions of matrices. For example, when the dominant eigenvalues are not relatively well separated, convergence is exceedingly slow and truncation could result in unpredictable numerical errors. Also, the time complexity of the existing algorithms indicates their impracticality for applications involving very large matrices. As a result, parallel and distributed algorithms have been regarded as the promising ground for discovery of new algorithms that could handle very large matrices, and numerous papers have taken the task of parallelizing deterministic algorithms. Subsequently, because of the lower time complexity of the approximation methods compared with direct methods, researchers moved toward parallelizing the approximation methods [6] [7] [13]. However, it is often the case that approximation methods fall into a local extremum of the error function, and convergence fails to reach the desired neighborhood of the true answer.

In this article, we introduce a new parallel algorithm based on iterative methods to extract eigenvalues and construct eigenspaces. We introduce a combination of simulated annealing and variants of evolutionary algorithms from the theory of combinatorial algorithms to improve the accuracy of our algorithm. We demonstrate that our method (called EPMP) greatly accelerates the rate of convergence. Test-case numerical examples confirm that computation times can be dramatically reduced, while accuracy is superior if comparable computation times are allocated. EPMP works most effectively on the genre of problems that pose numerical challenges when using the traditional power method. Another advantage of EPMP is its flexibility to generalize to other extensions of the eigenvalue problems and variants of the power method. We will briefly comment on a few selected generalizations of EPMP.

We found the Arnoldi iterative method [2] to be a good approximation strategy that lends itself well to parallelization in the steps based on matrix multiplication. A combination of simulated annealing and the theory of genetic algorithms is embedded into the iterative method to improve our method's accuracy. Before we embark on illustrating our methodology, the preliminaries are reviewed in Section 2, followed by illustration of the combinatorial method, the algorithm of the parallelized iterative method, and the combination of combinatorial and iterative methods in Section 3. Section 4 describes some preliminary results of our algorithm.

## 2 Preliminaries

This section outlines the basic concepts that will be used in our EPMP method.

### 2.1 Iterative Method

The Power method, for general square matrices, is the simplest of all the methods for solving for eigenvalues and eigenvectors. The basic idea is to multiply the matrix A repeatedly by a well-chosen starting vector so that the component of that vector in the direction of the eigenvector with largest eigenvalue (absolute value) is magnified relative to the other components. In theory, exact solutions could be reached in an infinite number of iterations; in practice, the process must stop after some finite number of iterations. Therefore, some auxiliary stop-time criteria must be added to iterative algorithms, so conditions are determined for sufficient approximation of the solution and to report the current solution as the best answer, as in [4] [10] [11].

Below is a version of the Arnoldi algorithm; see [12] for more information.
*Iterative Algorithm: The Arnoldi Iteration: The algorithm applied to a given matrix* A.
1. Start with an arbitrary vector $q_1$ with norm *1*.
2. Repeat for *k=2,3,…*
    $q_k \leftarrow Aq_{k-1}$
    for *j=1* to *k-1*
        $h_{j,k-1} \leftarrow q_j^* q_k$
        $q_k \leftarrow q_k - h_{j,k-1} q_j$
    $h_{k,k-1} \leftarrow \| q_k \|$
    $q_k \leftarrow q_k / h_{k,k-1}$

### 2.2 Simulated Annealing

The simulated annealing method is a popular method used in optimization to avoid the iteration steps getting trapped in a local optimum. Simulated annealing was inspired by a method for cooling metal. In each step of iteration, the algorithm selects a neighbor of the current solution from the feasible solution space and replaces the current solution with the neighbor when specified constraints are satisfied. For a randomly selected neighbor *Y* of the current solution *C*, if *Y* is more likely to be closer to the true solution, then the algorithm replaces *C* with *Y*, although the algorithm could also perform such replacements randomly. The probability of replacement is determined by the value of *temperature*. The *temperature* value is initialized by a random value $T_0$. In each iteration, it is reduced by a reduction rate *α* as a cooling schedule. By reducing the temperature value, the probability is also reduced. More exhaustive information can be found in [8].

### 2.3 Genetic Algorithms

Genetic Algorithms (GA) are a subclass of Evolutionary Algorithms (EA) with broad searching within the feasible solution space. The idea is to make unbiased random choices at the outset, and repeatedly analyze the choices in order to increase the probability of reaching the optimal solution. This kind of EA is inspired by population biology, where analysis of a population of solutions leads to selection strategies that evolve the progenies toward the optimal solution. A discussion of convergence and criteria for stopping time can be found in [3], and many articles have extended these results to a broad range of optimization problems. According to GA, competition among the solutions in selection for the next generation (i.e. iteration stage) is imposed according to a scoring function (the fitness function) that measures the optimality of the solutions in each generation. The competition is performed in some iteration to find the best score for global or even local optimality. In each step, the algorithm searches between the current population's neighbors to find new candidates for the optimal solutions. Selection of new neighbors is performed by applying two evolutionary methods: *mutation* and *crossover*. The mutation process is performed on a set of solutions by substituting some features of the solution members so that the new generation of solutions could gain higher score according to the fitness function, and consequently, provide better candidates than the root solution. The crossover process searches neighbors and, by crossing two solutions from the current population, creates two other solutions and increases the likelihood of improvement in fitness score.

These evolutionary processes attempt to create and check other members of the feasible solution space to reach better results and escape from local optimal solution. In [3] and [8], the algorithm is described in detail.

## 3  Method

This section focuses on the EPMP method, discussing first the combinatorial algorithms. As described in the previous section, *simulated annealing* attempts to escape from local extrema by using the dynamic *temperature* variable. In the case of the iterative method, there is not a set of feasible solutions from which simulated annealing theory could select another solution. Therefore, another method for generating new solutions is needed. The new solution must be close enough to current solution—to keep the feasibility—and must also, based on simulated annealing, be a different solution that could check other points of the solution space. Here, the *mutation function* could be helpful to generate a new solution. As mentioned, based on the predefined probability value, the mutation function substitutes the solution's features and generates new solutions from the current one. The following pseudo-code provides the steps of computing an eigenvector using a combinatorial algorithm.

```
1-   read Matrix dimension "n" from file.
2-   Read Matrix "A" from file.
3-   v = random vector.
4-   While stop condition not satisfied, do
     a.  v = v / length(v)
     b.  y = v.A
     c.  if abs((P(y)-P(v))/P(y))<eps then
         i.    v = y
         ii.   Stop condition satisfied.
               Break loop.
     d.  if P(y) < P(v) then
         v = y
     e.  else
         i.    r = rand
         ii.   if    r<exp((P(y)-P(v))/T)
               then Mutation(v)
         iii.  else
                  v = y
         iv.   T = T*alpha.
5-   lambda = length(v)
6-   v = v / length(v)
```

Next, we discuss the parallelization strategy. Matrix multiplication is one of the classic problems that can be solved efficiently on parallel and distributed platforms. Numerous parallel algorithms have been proposed for this [5] [9]. The iterative method is based on matrix multiplication, so it can be parallelized. The following algorithm shows our parallelized iterative method into which the combinatorial algorithm has been imported.

```
1-  for i=1 to p do in parallel
    a. read Matrix dimension "n" from file
    b. rows = n / p.
    c. read i'th part (from row (i-1)*rows
       to i*rows-1) of Matrix from file and
       store in "A".
2-  Processor 1:
    v = randomVector and broadcast v.
3-  While stop condition not satisfied do
    a. v = v / length(v)
    b. for i=1 to p do in parallel
        Processor 1:
         1.  for j=1 to rows do
             a.  y[j]=0.
             b.  for k=1 to n do
                 y[j]+=A[(i-1)*rows+j][k]*v[k]
    c. Processor 1:
       gather y from all processors.
    d. Processor 1:
       broadcast y to all processors.
    e. if abs((P(y)-P(v))/P(y))<eps then
       i.  v = y
       ii. stop condition is satisfied.
           Break the loop.
    f. if P(y)<P(v) then
       v = y.
    g. else
       i.   Processor 1:
            R=rand and broadcast r
       ii.  if r<exp((P(y)-P(v))/T) then
            Processor 1: mutate v and
            broadcast v.
       iii. else
            v = y
       iv.  T = T*alpha.
4-  lambda = length(v)
5-  v = v / length(v)
```

Matrix-vector multiplication follows the common parallelization theory based on the well-known idea of dividing the rows of a given matrix according to the number of processors (p); and all rows receive the vector. Thereafter, each processor multiplies its associated rows by the given vector. For a matrix $A_{n*n}$ and a vector $v_{n*1}$ each row $A_{(i,-)}$ should be multiplied by the vector's entries and sum of multiplications, replaced by the *i*-th row in the result vector. Therefore, for each row, 2*n* operations are needed. According to parallelization, each processor receives *n/p* rows, and performs 2*n* operations. As there are *p* processors, the parallel algorithm time complexity is equal to $O(p*(n/p)*2*n) = O(n^2)$, which is equal to the sequential time complexity. To summarize, the parallel algorithm is *cost optimal* [1]. But, as becomes clear below, for $p \geq n$, the algorithm empirically is not optimal.

## 4  Experimental Results

This section presents some results that were obtained in the context of applications to computational biology. The parallelized method is compared to our sequential method in Figure 1 below, including the application matrix dimensions and their run-time complexity.

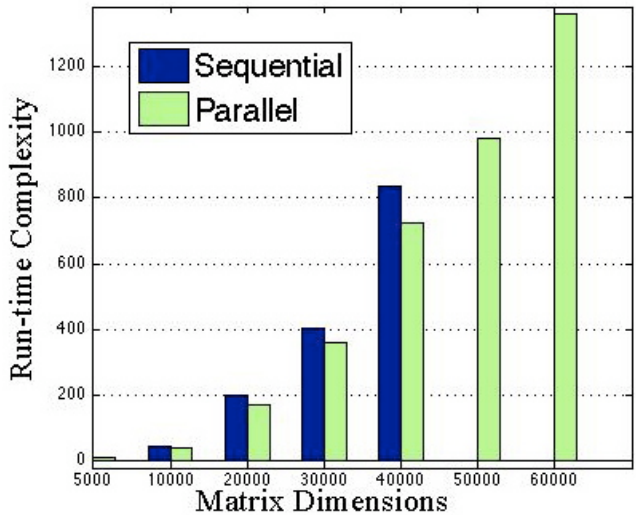

**Figure 1**: Comparison of sequential and parallel applications' running time complexity for different sized matrices. For each dimension, the average running time of 12 runs is indicated.

**Table 1**: Running times for the sequential and parallel applications for matrices of different sizes.

| Matrix Dimensions | Run-time Complexity, Sequential | Run-time Complexity, Parallel |
|---|---|---|
| 5000 | 12.108 | 11.116 |
| 10000 | 45.554 | 41.058 |
| 20000 | 200.66 | 173.206 |
| 30000 | 404.836 | 362.366 |
| 40000 | 834.432 | 722.176 |
| 50000 | NULL | 980.321 |
| 60000 | NULL | 1360.245 |

As shown in Table 1, sequential algorithms could not hand matrices with dimension more than 40000x40000 (and the same difficulty even with some matrices of smaller dimension.) Also, sequential algorithms take more time. For the comparisons, the sequential application ran on a Xeon quad 3 GHz with 8 GB memory, and Parallel algorithm main node used previous hardware and its three clients had a Xeon dual 2.4 GHz processor with 4 GB memory. The results for matrices of larger dimension require generalization of EPMP, and their discussion is postponed to a forthcoming article.

## 5 Conclusion

In this article, we proposed a novel, computationally efficient, and highly accurate approach for computation of matrix eigenvalues and eigenvectors. The approach stands on a parallel iterative method and uses a combination of simulated annealing and genetic algorithms to improve its accuracy. Parallelization of the algorithm makes it suitable for application to large matrices where intermediary combinatorial algorithms monitor the convergence, and to prevent the iteration sequence from getting trapped in a local extremum. Generalization of EPMP to extract additional eigenvalues and eigenvectors follows the deflation procedure introduced by Sorenson, where for nonsymmetric matrices, it is necessary to work with at least two vectors [18] [21]. The impact of the EPMP method is better appreciated when it is explicitly or implicitly part of more modern methods such as the QR method or the methods of Lanczos [21]. These extensions and other generalizations will appear in a forthcoming article.

## 6 Acknowledgements

We thank Professor Abbas Nowzari Dalini for his valuable contributions in the early stages of this work. We also wish to thank Meysam Bastani for his help during the implementation process.